\documentclass[10pt,aps,twocolumn,prc,superscriptaddress,showpacs,nofootinbib,noshowkeys,floatfix,preprintnumbers]{revtex4-1}
\usepackage{graphicx}
\usepackage[usenames]{color}
\usepackage{amsmath,amssymb}
\usepackage{multirow}
\usepackage{longtable}
\usepackage[normalem]{ulem}
\usepackage{epstopdf}
\usepackage{times}
\usepackage[normalem]{ulem}  

\renewcommand\sout{\bgroup \color{red} \ULdepth=-.5ex \ULset}
\usepackage{bm}
\graphicspath{{./Figs/}}
\begin{document}
\title{Size and Distance Scales in the Nucleon and in Dense Baryonic Matter}
\date{\today}
\author{Wolfram Weise}
\affiliation{Technical University of Munich,  School of Natural Sciences,  Physics Department, 85747 Garching, Germany}
\affiliation{Excellence Cluster ORIGINS, Boltzmannstr.\,2, 85748 Garching, Germany}

\begin{abstract}
This memorial tribute to Mannque Rho follows a line of thoughts and ideas that he continuously inspired and shaped over many decades: from the two-scales picture of low-energy nucleon structure to dense and cold baryonic matter as it is realized in the cores of neutron stars.  Early groundbreaking concepts are recalled and updated by recent advanced analyses of the `core' and `cloud' sizes of the nucleon.  Implications for dense nuclear matter are then discussed and confronted with empirical information from Bayesian inference analyses of neutron star observables.
\end{abstract}
\maketitle

\section{Prologue}
Mannque Rho's legacy is deeply interconnected with the outstanding role played by symmetry and topology in the physics of the Strong Interaction.  Over six decades of his continuously creative and productive scientific life he left his strong imprint on basically every aspect of chiral symmetry and its spontaneous breaking that governs the low-energy,  long-wavelength limit of QCD.  Here is a list of selected key headlines: Meson exchange currents and axial currents in nuclei \cite{Chemtob1971, Kubodera1978, Rho1991}; Baryons as chiral solitons and the Little Bag \cite{BR1979, Brown1979,Vento1980, Rho1983, Jackson1983}; Chiral restoration in dense/hot matter and Brown-Rho scaling \cite{BR1991, BR1996, BR2002}.  

The present author's own fruitful contacts with Mannque's work and his way of thinking,  often congenially inspired by the intuition of Gerry Brown during visits to Stony Brook,  began during the development period of the chiral bag model.  Spontaneously broken chiral symmetry implies that pions as (approximate) chiral Nambu--Goldstone bosons couple to the nucleon and add a `soft' surface degree of freedom to its structure.  This picture of the nucleon and attempts to quantify a delineation between its compact core and the meson cloud were topics of early joint publications \cite{BRW1986, BKRW1988}.  A decade later Mannque visited our group at TU Munich as a recipient of a Humboldt Research Award (Forschungspreis),  and we enjoyed exploring the effective mass of kaons in dense baryonic matter \cite{WRW1997}.  Still decades later we got together again and wrote a review article \cite{HRW2016} that surveyed work we were involved in from various points of view,  all guided by the principles of chiral symmetry and effective field theories for hadronic, nuclear and stellar matter.

This tribute to Mannque returns to themes related to the two-scales (core + cloud) description of the nucleon that he influenced and shaped for long periods of his scientific life.  In a recent publication \cite{Rho2024} he gave an illuminating account of the historical disputes confronting the early MIT bag model of the nucleon and its large bag radius,  $R\sim 1$ fm,  with the small chiral bag and the role of topology in the latter.  In either model the size scales associated with the distributions of baryon number and other charges inside the nucleon are key issues. The emergence of much improved nucleon form factor data and their detailed interpretation can nowadays help providing more quantitative conclusions about the delineation of the nucleon's `core'  and `cloud' sizes.  

At the same time such considerations obviously have an important impact on persisting questions about the properties of dense matter as it exists in the cores of neutron stars,  another one of Mannque's long-term topics of active interest.  Significant progress has been achieved in recent years,  setting constraints on the equation-of-state of highly compressed cold baryonic matter through observations and advanced data analysis.  Selected highlights of these developments will also be reported here. 

\section{Sizes of the nucleon}
\label{sec:sizes}

As mentioned,  models based on chiral symmetry describe the nucleon as a complex system characterized by two scales: a compact `hard' core\footnote{The idea of a compact core in the center of the nucleon with a size much smaller than the proton charge radius found additional support by deep-inelastic scattering data at HERA and $J/\Psi$ coherent scattering on nucleons and nuclei \cite{Caldwell2010}.  Evaluations of the nucleon size deduced from high-energy nucleus-nucleus cross sections have pointed in a similar direction \cite{Nijs2022}.} and a surrounding `soft' quark-antiquark cloud in which pions play a prominent role.  The chiral bag \cite{BR1979} and other chiral quark models such as the cloudy bag \cite{Thomas1984} were designed with this concept in mind~\cite{TW2001}.  

\begin{figure}
\centering
\includegraphics[width=7cm]{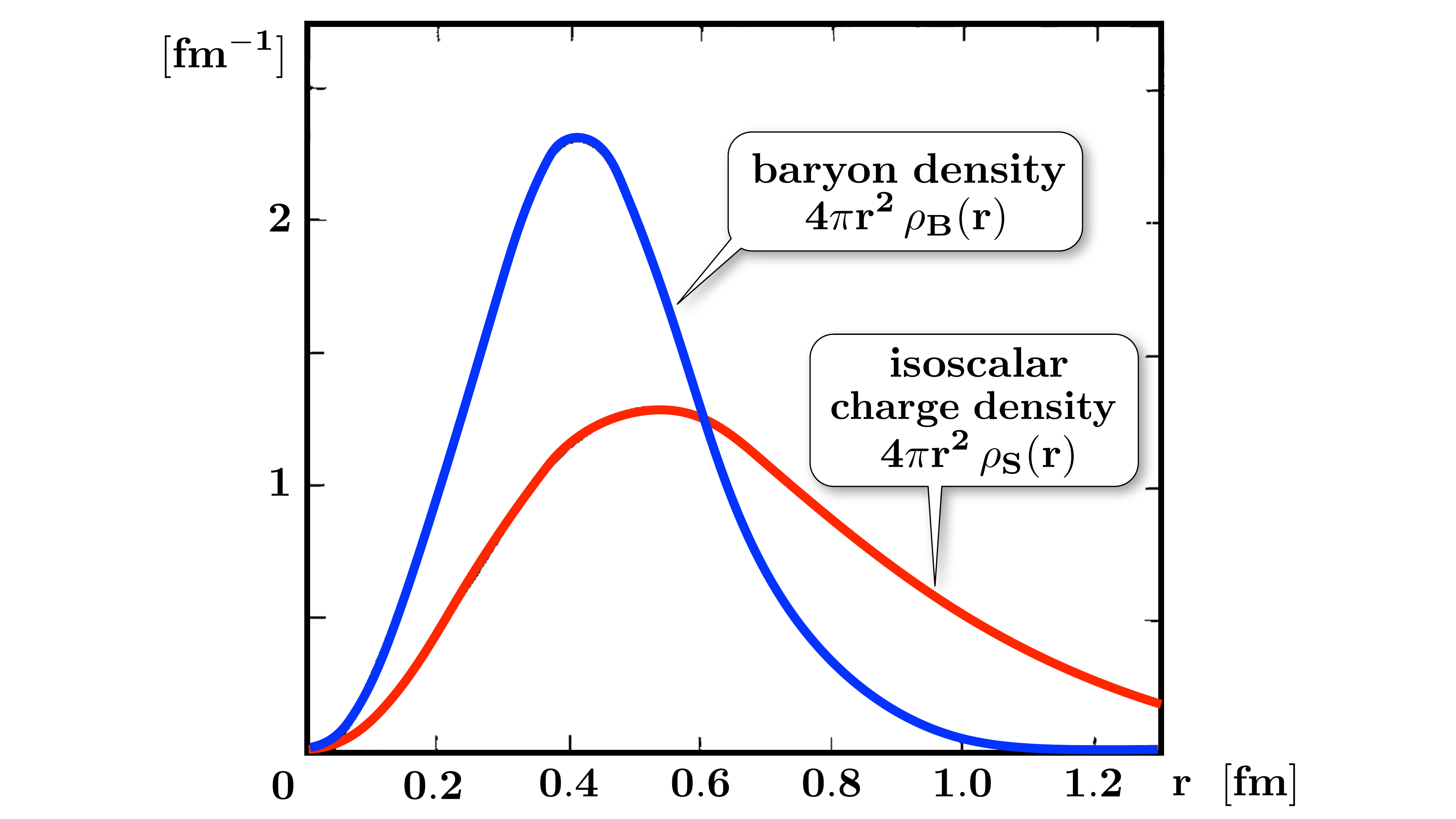}
\caption{Distributions of baryon number, $\rho_B(r)$,  and isoscalar electric charge, $\rho_S(r)$,  in the nucleon,  both multiplied by $4\pi r^2$.  Results of a topological chiral soliton (Skyrmion) model with vector mesons \cite{Meissner1987}.
\label{fig1} }
\end{figure}

Such a picture also emerged in descriptions of the nucleon as a chiral topological soliton (Skyrmion) with vector mesons \cite{Meissner1986, Meissner1987}.  The distinction between a compact distribution of baryon number and a more stretched distribution of isoscalar electric charge in the nucleon (see Figure\,\ref{fig1}) was an early manifestation of how the topological Skyrme soliton concept connected to the vector meson dominance principle \cite{HY2003}.  Mannque pursued and extended this basic idea in many ways.

The understanding of nucleon structure in the low-energy,  long-wavelength limit has in the meantime advanced to a level that enables a more quantitative evaluation of the core-plus-cloud scenario,  based on detailed analyses of nucleon form factors and radii \cite{KW2024}.  Primary empirical sources of information are the accurately determined isoscalar and isovector electric charge form factors.  Further input comes from the axial and the mass form factors.  The mean-squared radii associated with these measured form factors are all significantly different from one another,  indicating that there is no single and universal `size' scale of the nucleon.  However,  detailed spectral analyses have collected evidence for a common half-a-fermi sized core inside the nucleon which contains the three valence quarks and thus the baryon number distribution.  At the same time this core carries most of the nucleon mass generated by the (gluonic) QCD trace anomaly.  The mesonic clouds surrounding this core carry the quantum numbers of the currents governing the respective form factors.  These distinct mesonic surface components are shown to account for the observed differences in the empirical r.m.s.  radii.  

\subsection{Nucleon Form Factors and Radii}

Each given form factor $G(q^2)$ related to a current operator $J^\mu$ and its nucleon matrix elements, $\langle N(p')|J^\mu|N(p)\rangle$ $(q = p' - p)$,  has a representation in terms of an unsubtracted dispersion relation,  
\begin{eqnarray}
G(q^2) = \frac{1}{\pi}\int_{t_0}^\infty dt\, \frac{\text{Im} \,G(t)}{t-q^2-i\epsilon}~,\
\end{eqnarray} 
where $q^2=q_0^2 - \vec{q}^{\,2}$ is the squared four-momentum transfer.  The normalization $G(q^2= 0)$ is identified with the `charge' associated with the current $J^\mu$. 
Mean square radii are given as\footnote{Note that this Lorentz-covariant definition of $\langle r^2\rangle$ is independent of the choice of reference frame.  However,  an interpretation as a radius in three spatial dimensions is tied to a special coordinate system, the Breit frame with $p=(E,-\vec{\,q}/2), p'=(E,\vec{\,q}/2)$.}
\begin{eqnarray}
\langle r^2\rangle &=& \frac{6}{G(0)} \frac{dG(q^2)}{dq^2}\Big|_{q^2=0} \nonumber\\
&=& \frac{6}{\pi}\left[\int_{t_0}^{t_c}\frac{dt}{t^2}S(t)+\int_{t_c}^\infty\frac{dt}{t^2}S(t)\right]~,
\end{eqnarray}
where a delineation scale,  $t_c$ of order 1 GeV$^2$,  has been introduced.
The spectral distribution $S(t)= \text{Im}\,G(t)/G(0)$ includes intermediate hadronic states through which the external probing field couples to the respective nucleon current.  The low-$t$ range of this distribution ($t \lesssim t_c$) is expected to be associated with the mesonic surface,  while the high-$t$ region ($t > t_c$) supposedly reflects the nucleon core.   Three different form factors (electric,  axial and mass form factors) are of special interest in this context.

\subsubsection{Isoscalar and isovector electric form factors and radii}

The nucleon matrix elements $(N=p,n)$ of the electromagnetic current,
\begin{eqnarray}
& &\langle N(p')|J^\mu_{em}|N(p)\rangle = \nonumber \\ 
& &\bar{u}(p')\left[F_1(q^2)\gamma^\mu+{i\over 2M}F_2(q^2)\sigma^{\mu\nu}q_\nu\right]u(p)~,
\end{eqnarray}
define the Dirac and Pauli form factors,  $F_1(q^2)$ and $F_2(q^2)$.  The four-momentum transfer is $q^\mu = (p' - p)^\mu$,  and $M$ denotes the nucleon mass.  The proton and neutron electric form  factors are given by:
\begin{eqnarray}
 G_E^{p,n}(q^2)= F_1^{p,n}(q^2)+ {q^2\over 4M^2}F_2^{p,n}(q^2)~,
\end{eqnarray}
with charges $G_E^{p}(0) = 1$ and  $G_E^{n}(0) = 0$.  The isoscalar and isovector combinations,
\begin{eqnarray}
 G_E^{S,V}(q^2)= {1\over 2}\left[G_E^{p}(q^2)\pm G_E^{n}(q^2)\right]~,
\label{eq:GE}
\end{eqnarray}
are of special interest here.

The slopes of $ G_E^{p,n}$ at zero momentum transfer determine the corresponding mean-squared radii.  The empirical r.m.s.  proton charge radius has been obtained in electron scattering and muonic hydrogen measurements \cite{Pohl2010} reviewed in \cite{Gao2022} and consistently updated in \cite{Lin2021, Lin2022}: $\langle r_p^2\rangle^{1/2} = 0.840\pm 0.003\pm 0.002$ fm. Its combination with six times the slope of the neutron electric form factor,  $\langle r_n^2\rangle = -0.105\pm 0.006$ fm$^2$ \cite{Filin2021},  gives the isoscalar and isovector mean-squared charge radii of the nucleon,  $\langle r^2_{S,V}\rangle = \langle r_p^2\rangle \pm \langle r_n^2\rangle$,  resulting in the following values:
\begin{eqnarray}
\sqrt{\langle r_S^2\rangle} = 0.78 \pm 0.01 \,\text{fm}~, ~\sqrt{\langle r_V^2\rangle} = 0.90\pm 0.01 \,\text{fm}.
\label{eq:radii}
\end{eqnarray}
Advanced lattice QCD simulations \cite{Djukanovic2024} have reached a level of precision that closely approaches these empirical radii.  

The combination of isoscalar and isovector electric form factors is a suitable set for discussing a delineation between the `core' and `cloud' parts of the nucleon.  We start with the isoscalar form factor and write it again as an unsubtracted dispersion relation:
\begin{eqnarray}
G_E^S(q^2)=\frac{1}{\pi}\int_{t_0}^\infty dt\, \frac{\text{Im} \,G_E^S(t)}{t-q^2-i\epsilon}~,
\label{eq:DR}
\end{eqnarray} 
normalized as $G_E^S(0) = {1\over 2}$. The spectrum $\text{Im} \,G_E^S(t) = \text{Im} \,F_1^S(t)+{t\over 4M^2}\,\text{Im} \,F_2^S(t)$ with $F_i^S = {1\over 2}(F_i^p + F_i^n)$ starts at the three-pion threshold, $t_0 = 9m_\pi^2$.  It is strongly dominated by the narrow $\omega$ meson while the contribution of the isoscalar $3\pi$ continuum in the range $t\le m_\omega^2$ is negligibly small~\cite{Kaiser2019}.  Additional contributions come from the $\phi$ meson,  its $K\bar{K}$ tail and the $\rho\pi$ continuum. 

A quick first estimate can be obtained using the simplest version of a vector meson dominance model (VDM). In this model the probing isoscalar $J^P=1^-$ photon converts into an omega meson which couples to the nucleon core.  The flavour SU(3) Gell-Mann - Nishijima formula,  $Q = I_3 + {1\over 2}(B+S)$,  relates the isoscalar charge $Q = {1\over 2}$ to the baryon number $B=1$ for $I_3=S=0$.  The distribution of baryon number carried by the three valence quarks in the nucleon core can therefore be identified with the core part of the isoscalar charge distribution.  The surrounding quark-antiquark cloud represented by the $\omega$ meson does not contribute to baryon number and electric charge but adds to determining the isoscalar radius, $\sqrt{\langle r_S^2\rangle}$.  In this picture the isoscalar electric form factor is given by the following ansatz:
\begin{eqnarray}
G_E^S(q^2) = \frac{F_B(q^2)}{2(1+|q^2|/m_\omega^2)}~.
\end{eqnarray}
The form factor $F_B(q^2)$ of the baryon number distribution in the nucleon core (with $F_B(0)=B=1$) acts as the source of the $\omega$ field that propagates with its mass $m_\omega$.  Introducing the mean-squared radius of the baryon core,  $\langle r_B^2\rangle = 6\frac{dF_B(q^2)}{dq^2}\big{|}_{q^2=0}$,  the mean-squared isoscalar charge radius becomes
\begin{eqnarray}
\langle r_S^2\rangle = \langle r_B^2\rangle + \frac{6}{m_\omega^2}~.
\end{eqnarray}
Using $m_\omega = 783$ MeV and the empirical value (\ref{eq:radii}) for $\langle r_S^2\rangle^{1/2}$,  the estimated core radius is 
\begin{eqnarray}
\langle r^2_S\rangle^{1/2}_{\text{core}} \equiv \sqrt{\langle r_B^2\rangle} \simeq  0.48\pm 0.01\,\text{fm}~.
\label{eq:VDMradius}
\end{eqnarray}

A core size of about 1/2 fm is indeed characteristic of chiral `core + cloud' models of the nucleon.  As shown in \cite{KW2024} this property is retained in a more detailed and realistic treatment of the spectral distributions governing the isoscalar form factor.  The analysis makes use of the precision fits 
 to $G_E^S(q^2)$ performed in \cite{Lin2022} for both spacelike and timelike regions of $q^2 = q_0^2 - \vec{q}\,^2$,  starting from 
\begin{eqnarray}
& &G_E^S(q^2)=  \nonumber \\ 
& &\frac{1}{\pi}\int_{9m_\pi^2}^\infty {dt\over t-q^2-i\epsilon}
\left[\text{Im}\,F_1^S(t) + {t\over 4M^2}\text{Im}\,F_2^S(t)\right]~.
\label{eq:DR2}
\end{eqnarray} 
The isoscalar mean-squared radius is:
\begin{eqnarray}
\langle r_S^2\rangle = {12\over\pi} \int_{9m_\pi^2}^\infty dt\left[{\text{Im}\,F_1^S(t)\over t^2} + {\text{Im}\,F_2^S(t)\over 4M^2 t}\right]~.
\label{eq:msradius}
\end{eqnarray} 
\begin{figure}
\centering
\includegraphics[width=6cm]{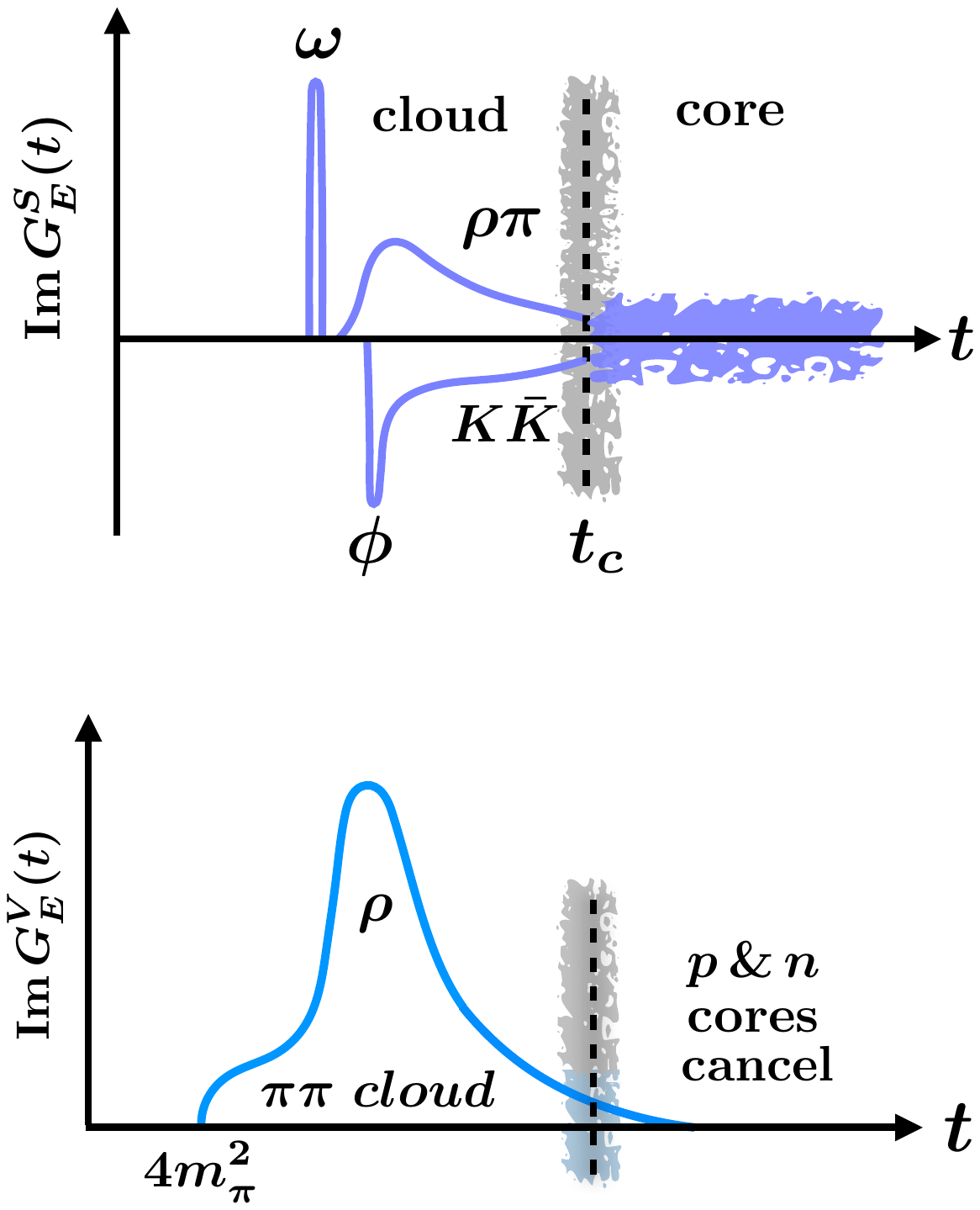}
\caption{Schematic spectral functions of mesonic cloud contributions to the isoscalar and isovector electric form factors of the nucleon.  Upper (isoscalar) section shows $\omega$ and $\phi$ mesons together with $\rho\pi$ and $K\bar{K}$ continuum parts.  The high-mass region $t > t_c$ stands for the nucleon core (figure partly adapted from \cite{Meissner2023}).  Lower (isovector) section shows $\rho$ meson and $\pi\pi$ contributions while proton and neutron cores (almost) cancel.
\label{fig2} }
\end{figure}
The fits to the spectral functions,  $\text{Im}\,F_{1,2}^S(t)$,  include the $\omega$ and $\phi$ meson poles together with $\rho\pi$ and $K\bar{K}$ continuum parts as sketched in Figure\,\ref{fig2}.  These mesonic contributions cover a range $t \lesssim t_c$,  associated with the meson cloud.  The short-distance core part refers to the region $t > t_c$.  It includes the information from the timelike domain measured in $e^+e^-\rightarrow N\bar{N}$.  This process directly involves the valence quark core of the nucleon (and of the antinucleon) through their production.  The timelike sector of the form factor is thus particularly informative about properties  of the core.  In practice this region is parametrized by a series of high-mass poles \cite{Lin2022}. 

The detailed analysis \cite{KW2024} of the spectral distributions separated into low-mass mesonic parts and high-mass poles gives the following result for the mean-squared isoscalar core radius,  $\langle r^2_S\rangle_{\text{core}} =  \langle r^2_S\rangle - \langle r^2_S\rangle_{\text{mesonic}}$:
\begin{eqnarray}
 \langle r^2_S\rangle_{\text{core}} \simeq (0.23 + 0.02)\,\text{fm}^2~,
\label{eq:msradius3}
\end{eqnarray} 
where the leading number in brackets comes from $\text{Im}\,F_1^S$ while the smaller piece refers to $\text{Im}\,F_2^S$.  
With inclusion of a (conservative) uncertainty estimate,  
\begin{eqnarray}
\langle r_S^2\rangle^{1/2}_{\text{core}} \simeq 0.50\pm 0.01\,\text{fm}~,
\label{eq:DR3}
\end{eqnarray} 
turns out to be remarkably close to the simplest VDM estimate (\ref{eq:VDMradius}).  

Of course,  conclusions drawn about the core radius are tied to the choice of the delineation scale, $t_c$,  that separates low- and high-mass sectors of the spectral distributions under the integral in (\ref{eq:msradius}) and may thus appear to be highly model dependent.  

At this point the {\it isovector} form factor, $G_E^V(q^2)$,  enters as an important test object in an analogous spectral analysis.  It involves the {\it difference} of proton and neutron form factors in (\ref{eq:GE}).  In the limit of perfect isospin symmetry the $uud$ and $udd$ valence quark cores of proton and neutron are identical.  A first guess therefore leads to expect that these cores cancel in $G_E^V$,  i.e.  the isovector {\it core} radius should vanish: $\langle r_V^2\rangle_{\text{core}} =0$.  A grossly incorrect choice of $t_c$ would presumably imply a substantial deviation from this limit.  The inspection of the isovector core radius using again the corresponding fit  in \cite{Lin2022} yields the result \cite{KW2024}:
\begin{eqnarray}
 |\langle r^2_V\rangle_{\text{core}}| \simeq 0.02 \,\text{fm}^2 \,.
\label{eq:msradius4}
\end{eqnarray} 
(A small deviation from zero is expected because of isospin breaking effects).  Hence the observed cancellation of the proton and neutron `core' parts in the isovector form factor is in essence an indirect confirmation of the two-scales core-plus-cloud structure seen in the analysis of the isoscalar charge radius.

The isovector charge radius of the nucleon thus arises almost entirely from the interacting two-pion cloud  governed by the $\rho$ meson and the prominently enhanced low-mass tail that extends down to the $\pi\pi$ threshold,  $t_0 = 4m_\pi^2$,  as illustrated in Fig.\,\ref{fig2}.  It is interesting to note that a recent reanalysis of isovector electromagnetic form factors including a detailed explicit treatment of the two-pion continuum \cite{KLM2026} also confirms a `core' size of about 1/2 fm.

\subsubsection{Axial form factor and radius}

Another object of interest is the form factor $G_A(q^2)$ associated with the axial vector current of the nucleon.  It has been deduced \cite{Hill2018} from weak muon capture on the proton,  $\mu^-\,p \rightarrow \nu_\mu \,n$,  from neutrino scattering on the deuteron and from pion electroproduction,  $e\,N \rightarrow e\,N'\,\pi$.  

The low-$q^2$ expansion of the axial form factor determining the mean-square radius $\langle r_A^2\rangle$,
\begin{eqnarray}
G_A(q^2) = G_A(0)\left[1 + {1\over 6}\langle r^2_A\rangle q^2 + \dots\right]~,
\label{eq:axialff}
\end{eqnarray} 
involves the axial vector coupling constant, $g_A = G_A(0)$.  From neutron beta decay,  $g_A = 1.2764(8)$ \cite{Maerkisch2019}.  \footnote{The extraction from pion electroproduction makes use of the Goldberger-Treiman (GT) relation,  $g_A = g_{\pi NN}\,f_\pi/M_n$.  With the pion-nucleon coupling constant $g_{\pi NN} = 13.1$,  the pion decay constant $f_\pi = 92.3$ MeV and the neutron mass $M_n = 939.6$ MeV,  the resulting $g_A^{GT}$ differs from the empirical $g_A$ by less than 1\%. } 

Determinations of $\langle r_A^2\rangle$ reported in \cite{Hill2018} refer to two sources of information: a combined dipole fit to the axial form factor extracted from $\nu d$ scattering and pion electroproduction, which gives $\langle r_A^2\rangle = 0.454\pm 0.013$ fm$^2$,  and a more conservative analysis of $\nu d$ scattering and $\mu p$ capture data, without resorting to an assumed dipole form,  which consequently involves larger uncertainties: $\langle r_A^2\rangle = 0.46\pm 0.16$ fm$^2$.
In either of these two cases the axial radius is evidently much smaller than the proton charge radius.

Writing the axial formfactor as an unsubtracted dispersion relation,
\begin{eqnarray}
G_A(q^2)= \frac{1}{\pi}\int_{t_0}^\infty dt\, \frac{\text{Im}\,G_A(t)}{t-q^2-i\epsilon}~,
\label{eq:axialDR}
\end{eqnarray}
and recalling the normalisation $G_A(q^2=0)= g_A$,  the corresponding mean-squared radius is:
\begin{eqnarray}
\langle r_A^2\rangle &=& \frac{6}{g_A} \frac{dG_A(q^2)}{dq^2}\Big|_{q^2=0}\nonumber\\
&=&\frac{6}{g_A \pi}\int_{t_0}^\infty\frac{dt}{t^2}{\text{Im}\,G_A(t)}~,
\label{eq:Raxial}
\end{eqnarray}
The isovector $J^P=1^+$ spectrum,  Im\,$G_A(t)$,  starts at the three-pion threshold, $t_0=9m_\pi^2$,  and prominently features the broad $a_1$ meson resonance.  An approximate scale of this `cloud' part can be estimated by introducing an $a_1$ pole with a mass $m_a\simeq 1.2$ GeV.  Using the empirical $\langle r_A^2\rangle$ one finds for the remaining `core' size:
\begin{eqnarray}
\langle r_A^2\rangle_\text{core}^{1/2} = \left(\langle r_A^2\rangle - {6\over m_a^2}\right)^{1/2} 
\simeq 0.54\pm 0.02\,\text{fm} ,
\end{eqnarray}
if the dipole fit value of $\langle r_A^2\rangle$ is taken for reference.  Using instead the `unbiased' fit value the uncertainty in $\langle r_A^2\rangle_\text{core}^{1/2}$ increases to about 25\%.  

A detailed evaluation requires full account of the broad isovector $J^P=1^+$ three-pion spectral distribution.  A suitable ansatz is:  
\begin{eqnarray}
G_A(q^2) = {g_A\,m_a^2\over m_a^2-q^2+ \Sigma_a(q^2) -\text{i}\,m_a\,\Gamma_a^2(q^2)}~.
\end{eqnarray}
The self-energy correction $\Sigma_a(q^2)$,  compatible with the dispersion relation (\ref{eq:axialDR}),  is determined by a twice-subtracted dispersion relation:
\begin{eqnarray}
\Sigma_a(q^2) = {q^2\over\pi}(q^2-m_a^2)~{\cal P}\int_{9m_\pi^2}^\infty {dt\over t}{m_a\,\Gamma(t)\over (t-m_a^2)(t-q^2)}~.\nonumber\\
\label{eq:selfE}
\end{eqnarray}
Results from $\tau\rightarrow \pi\pi\pi\nu_\tau$ decays are used to set constraints on the energy dependence of the $a_1$ width,  $\Gamma_a(t)$.  The computation including this width yields \cite{KW2024}: 
\begin{eqnarray}
\langle r_A^2\rangle_\text{core}^{1/2} \simeq 0.53\pm 0.02\,\text{fm}~, 
\label{eq:RAcore}
\end{eqnarray}
with an estimated uncertainty based on the dipole fit to the empirical form factor.  A correspondingly larger uncertainty results if the unconstrained fit is used.  

The core radius (\ref{eq:RAcore}) deduced from the axial form factor is less accurately determined than the core radius (\ref{eq:DR3}) resulting from the analysis of the isoscalar electric form factor.  It is nonetheless remarkable that,  starting from two independent form factors with quite different empirical radii,  one consistently arrives at a common half-fermi scale for the core size in the nucleon.  

\subsubsection{Mass form factor and radius}
 
\begin{figure}
\centering
\includegraphics[width=8cm]{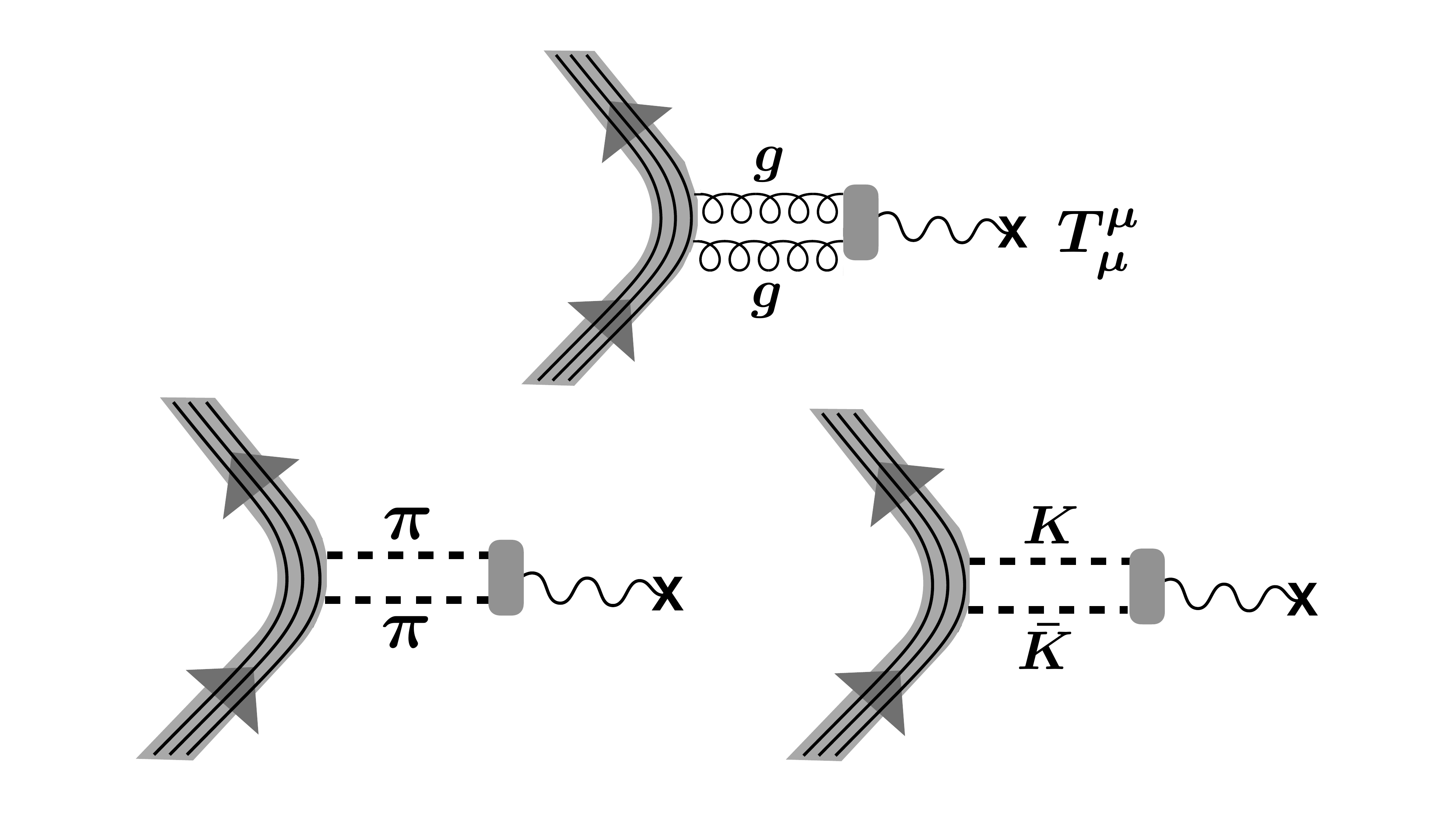}
\caption{Illustration of contributions to the spectrum,  $\text{Im}\,G_m(t)$,  of the nucleon's mass form factor: leading gluonic component (upper diagram); $\pi\pi$ and $K\bar{K}$ contributions (lower diagrams).
\label{fig3} }
\end{figure}

A further interesting quantity in this context is the mass radius of the proton deduced from $J/\psi$ photoproduction data~\cite{Caldwell2010, Kharzeev2021}.  The $c\bar{c}$ pair that forms the $J/\psi$ acts as a small dipole that couples to the nucleon through leading two-gluon exchange in QCD.  As demonstrated in \cite{Kharzeev2021} the amplitude for this process close to $J/\psi$ production threshold is proportional to the matrix element of the trace of the QCD energy-momentum tensor,  $T_\mu^\mu$.  With proper normalization this matrix element is referred to as the `mass' (or `gravitational') form factor of the nucleon:
\begin{eqnarray}
G_m(q^2) &=& \langle N(p')|T_\mu^\mu|N(p)\rangle \nonumber\\
&=& G_m^{(0)}(q^2) + \sigma_{N}(q^2) + \sigma_s(q^2)~.
\label{eq:Gmass}
\end{eqnarray}
The first term in (\ref{eq:Gmass}) is the gluonic form factor,
\begin{eqnarray}
G_m^{(0)}(q^2) = \langle N(p')|{\beta\over 2g}G_{\mu\nu}^a G^{\mu\nu a}|N(p)\rangle~,
\end{eqnarray}
where $\beta = -{bg^3\over 16\pi^2}$ is the QCD beta function with $b=11-2N_f/3=9$ for $N_f=3$ light quark flavors\footnote{Heavy ($c, b$ and $t$) quarks appear only as virtual $Q\bar{Q}$ loops in gluon propagators.  Their mass terms in $T_\mu^\mu$ cancel against corresponding heavy-quark sectors in the gluon term.}.  The scalar form factors, 
\begin{eqnarray}
\sigma_{N}(q^2) &=& \langle N(p')|m_l(\bar{u}u +\bar{d}d)|N(p)\rangle~, \\
\sigma_s(q^2) &=& \langle N(p')|m_s\bar{s}s|N(p')\rangle~,
\label{eq:sigmaff}
\end{eqnarray}
include small quark mass contributions,  with the average of the $u$- and $d$-quark masses,  $m_l = {1\over 2}(m_u + m_d)$,  and the strange quark mass $m_s$.  Altogether these terms represent the pieces illustrated in Figure\,\ref{fig3} from gluon-dominated short-distance structures,   $\pi\pi$ and $K\bar{K}$ continuum contributions,  respectively.

A once-subtracted dispersion relation representation of the mass form factor,
\begin{eqnarray}
G_m(q^2) = M +{q^2\over\pi} \int_{t_0}^\infty{\text{Im}\,G_m(t)\over t(t-q^2 - i\epsilon)}~,
\label{eq:Gm}
\end{eqnarray}
displays the normalisation to the nucleon mass,  $G_m(0) = M$.  The quark mass contributions are given by the pion-nucleon and strangeness sigma terms,  
\begin{eqnarray}
\sigma_{N}\equiv  \sigma_{N}(q^2=0)\quad\text{and}\quad\sigma_s\equiv \sigma_s(q^2=0)~.
\label{eq:sigma}
\end{eqnarray}
In the overall sum, 
\begin{eqnarray}
M = M_0 + \sigma_{N} + \sigma_s~,
\end{eqnarray}
the dominant piece $M_0$ refers to the `core' mass generated by the gluonic trace anomaly (the gluonic terms in $T_\mu^\mu$),  while the sigma terms account together for less than 10\% of the total $M$. 

The $J^P = 0^+$ two-gluon system couples strongly to the scalar-isoscalar two-pion continuum.  The lower limit in the spectral integral (\ref{eq:Gm}) is therefore at $t_0 = 4m_\pi^2$.  Unlike the prominent low-mass $\pi\pi$ spectrum with $J^P = 1^-$ in the isovector electric form factor,  the scalar-isoscalar $\pi\pi$ continuum contribution to $G_m(q^2)$ is however suppressed by the small ratio $\sigma_{N}/M$.  

The mean-squared radius associated with $G_m(q^2)$,
\begin{eqnarray}
\langle r_m^2\rangle = \frac{6}{M} \frac{dG_m(q^2)}{dq^2}\Big|_{q^2=0}=
\frac{6}{M \pi}\int_{t_0}^\infty\frac{dt}{t^2}{\text{Im}\,G_m(t)}~,\nonumber\\
\label{eq:Rmass}
\end{eqnarray}
has been extracted from differential $J/\psi$ photoproduction cross section measurements by GlueX at JLab.  The result quoted in~\cite{Kharzeev2021},  $\langle r^2_m \rangle^{1/2} =  0.55\pm 0.03\,$fm,  is based on a dipole fit to $d\sigma(\gamma p\rightarrow \psi p)/dq^2$.  A more recent update \cite{GlueX2023} gives $\langle r^2_m \rangle^{1/2} =  0.53\pm 0.04\,$fm,  compatible with the previous result \footnote{This latter value of $\langle r^2_m \rangle^{1/2}$ represents an average from fits taken over a photon energy range $E\gamma =8.9 - 10.8$ GeV.  The quoted uncertainties may be underestimated because of the model dependence implied by assuming dipole forms in those fits.}.

The elements of the spectral distribution illustrated in Figure\,\ref{fig3},  namely short-distance two-gluon exchange plus longer range $\pi\pi$ and $K\bar{K}$ components,  imply the following decomposition:
\begin{eqnarray}
\langle r^2_m \rangle= {M_0\over M}\langle r^2_0 \rangle+{\sigma_{N}\over M}\langle r^2_{\pi\pi} \rangle+{\sigma_s\over M}\langle r^2_{K\bar{K}} \rangle~.
\label{eq:Rm2}
\end{eqnarray}
\\
The dominant gluonic trace anomaly contribution with mass $M_0$ is identified with the squared `core' radius,  $\langle r^2_0 \rangle \equiv \langle r^2_m \rangle_{\text{core}}$, while the small corrections from $\pi\pi$ and $K\bar{K}$ `cloud' pieces involve the sigma terms (\ref{eq:sigma}).  
So far there is not yet a fully consistent picture for the sigma term $\sigma_{N}$.  Lattice QCD computations \cite{Agadjanov2023} give $\sigma_{N} = 43.7\pm 1.2\pm 3.4$ MeV,  a value close to the result obtained decades ago in the time-honored work of \cite{Gasser1991}: $\sigma_{N} = 45\pm 8$ MeV.  Further studies \cite{Hoferichter2023} based on an updated assessment of pion-nucleon scattering data raised this sigma term to $\sigma_{N} = 55.9\pm 3.5$ MeV,  a value that is consistent with an extraction of $\sigma_N$ from pionic atom data \cite{FG2019}.

The original analysis \cite{Gasser1991} proposed a large radius of the isoscalar s-wave $\pi\pi$ distribution at the nucleon surface: $\langle r_{\pi\pi}^2\rangle^{1/2}\simeq 1.3$ fm.  Recent advanced evaluations \cite{Cao2025} suggest a smaller value,  $\langle r_{\pi\pi}^2\rangle^{1/2}\sim 1$ fm.

Following the inspection and discussion in \cite{KW2024},  the strange quark contribution to the mass radius,  the one involving the strangeness sigma term $\sigma_s$,  turns out to play only a very minor role.  Using in (\ref{eq:Rm2}) the values $\sigma_{N} \simeq 56$ MeV and $\langle r^2_{\pi\pi} \rangle \sim 1$ fm$^2$ together with $\langle r^2_m \rangle^{1/2} \simeq 0.54$ fm,  the radius of the compact gluonic core of the nucleon that contains most of its mass becomes:
\begin{eqnarray}
\langle r^2_m \rangle_{\text{core}}^{1/2}\equiv\langle r^2_0 \rangle^{1/2} =  0.51\pm 0.05\,\text{fm}~,
\label{eq:Rmcore}
\end{eqnarray}
where an (arguably still underestimated) 10\% error includes the combined uncertainty effects from empirical sources and sigma terms.  While these uncertainties are evidently much larger than those quoted in the extraction of the core radius (\ref{eq:DR3}) from the accurate isoscalar electric form factor data,  the general trend towards a common`1/2-fermi rule' for the nucleon core size is maintained. 

In summary,  the explorations of three different form factors consistently support a picture of the nucleon as a compact `hard' core with a radius of about 1/2 fm, surrounded by a `soft' surface of quark-antiquark pairs forming mesonic clouds.  The core hosts the three valence quarks with their baryon number.  It also contains most of the nucleon mass generated by gluon dynamics through the QCD trace anomaly.  At the same time the localisation of the (almost massless) valence quarks within the compact core volume implies spontaneously broken chiral symmetry.

Given the prominent role of gluons generating the mass accumulated in the central core of the nucleon,  a comparison with the size of a pure-glue system,  the $J^{PC} = 0^{++}$ glueball,  is instructive.  Lattice QCD computations of gravitational form factors of the scalar glueball \cite{Abbott2026} yield evidence that the glueball mass radius is very small,  even smaller than the core size in the nucleon: $\langle r^2_G\rangle^{1/2} \simeq 0.26\pm 0.03$ fm.  With the commonly accepted scalar glueball mass of 1.6-1.7 GeV \cite{Morningstar2025}, it is quite conceivable that an injection of three light valence quarks into this compact chunk of gluonic energy density can establish the mass and size scale of the nucleon core.

\section{Dense Baryonic Matter and Neutron Stars}

With a baryonic core size 
\begin{eqnarray}
R_\text{core} \simeq \langle r_S^2\rangle_{\text{core}}^{1/2}\simeq\langle r_A^2\rangle_{\text{core}}^{1/2} \simeq \langle r_m^2\rangle_{\text{core}}^{1/2} \sim {1\over 2}\,\text{fm}~,
\label{eq:Rmcore}
\end{eqnarray}
and a meson cloud elongation of typically $R_\text{cloud} \sim 1$ fm,  there is a significant separation of these characteristic volume scales for a nucleon in vacuum: $(R_\text{cloud}/R_\text{core})^3 \gg 1$.  This scale separation is expected to increase further in dense baryonic matter,  for the following reasons.
The properties of the soft multi-pion cloud are closely tied to spontaneously broken chiral symmetry and the approximate Nambu-Goldstone boson nature of the pion.  The size of this cloud is expected to increase with baryon density $n_B$,  along with a decreasing in-medium pion decay constant,  $f_\pi^*(n_B)$,  which acts as a chiral order parameter.  The baryonic core,  on the other hand,  is governed by gluon dynamics with no leading connection to chiral symmetry in QCD.  This core is therefore assumed to be quite stable against substantial changes with increasing density\footnote{This expected tendency is underlined e.g.  by computations using an advanced three-flavor Nambu - Jona-Lasinio type model \cite{Bentz2025}.  Baryon cores are treated as quark-diquark bound clusters which in turn act as sources for mesons treated as quark-antiquark modes.  In this model a nucleon core radius of 0.47 fm in vacuum increases by less than 10\% in nuclear matter at saturation density,  $n_0 = 0.16$ fm$^{-3}$.  Even at $n_B \sim 3\,n_0$ this increase does not exceed 15\%.},  up until the compact hard cores begin to touch and finally overlap.

Before entering a more detailed discussion of possible implications for compressed baryonic matter from such geometrical perspectives,  it is useful to summarize our current understanding of the equation-of-state based on the analysis of neutron star data. 

\subsection{Constraints on the equation-of-state of neutron star matter}

Much progress has been made in recent years collecting data for masses and radii of neutron stars.  Masses,  in particular those of the heavy two-solar-mass stars,  have been established by Shapiro delay measurements in binaries with white dwarfs as neutron star companions.  An exceptional case is the heaviest known ($\sim 2.3\,M_\odot$) and fast rotating galactic pulsar reported by the Keck observatory.  A vital role is played by the NICER observatory at the ISS,  detecting X-rays from hot spots at the surfaces of rotating neutron stars.  These data permit to set limits on neutron star radii,  analyses that are being progressively improved.  Together with other mass determinations,  the NICER masses and radii shown in Table\,\ref{tab:Nstardata} refer to latest updated values.  In addition to the data listed in the Table,  information about tidal deformabilities deduced from gravitational wave signals of neutron star mergers (GW170817) is also included. 
\begin{table}
\renewcommand{\arraystretch}{1.5}
\setlength{\tabcolsep}{10pt}
\begin{center}
\caption{Data base of neutron star properties (masses and radii) used in Bayes inference analyses of the speed of sound and equation of state of neutron star matter. }
\label{tab:Nstardata}
\begin{tabular}{|c||c|c|c|}  
\hline 
 PSR & Mass $M/M_\odot$ & Radius $R~[\mathrm{km}] $ & Ref.\\ \hline
J0348+0432 & $1.806\pm 0.037$ & -- & \cite{Saffer2025}\\
J1614-2230 & $1.937\pm 0.014$ & -- & \cite{Agazie2023}\\ 
J0952-0607& $2.32\pm 0.11$ & -- & \cite{Romani2025}\\
J0740+6620 & $2.073\pm 0.069$ & $12.76^{+1.49}_{-1.02}$ & \cite{Dittmann2024}\\
J0030+0451 & $1.43^{+0.20}_{-0.17}$ & $12.68^{+1.31}_{-1.04}$ & \cite{Kini2026}\\
J0614-3329 & $1.44^{+0.06}_{-0.07}$ & $10.29^{+1.01}_{-0.86}$ & \cite{Mauviard2025}\\ 
J0437-4715 & $1.418\pm 0.044$ & $13.45\pm 1.65$ & \cite{Miller2026}\\
\hline
\end{tabular}
\end{center}
\end{table}

These data are used as input in Bayes inference approaches with the aim of establishing posterior bands of empirically constrained equations-of-state (EoS) for neutron star matter.  This is a prime source of information on highly compressed,  strongly interacting matter at zero temperature. 

A key quantity is the speed of sound or its square,
\begin{eqnarray}
c_s^2= {\partial P(\varepsilon)\over\partial\varepsilon}~,
\label{eq:soundspeed}
\end{eqnarray}
the derivative of pressure, $P$,  with respect to energy density, $\varepsilon$.
The sound speed is particularly informative about the possible occurance of a phase transitions.  For example,  a first-order phase transition with Maxwell construction would manifest itself as a region of constant pressure in the EoS,  $P(\varepsilon)$,  so the speed of sound would drop abruptly to zero at the onset of the phase coexistence region and recover at its end point.  A continuous crossover,  while indicating itself just by a change of slope in $P(\varepsilon)$,  could still show up as a pronounced maximum in $c_s^2(\varepsilon)$.  It is thus useful to conduct the Bayesian inference procedure starting with the speed of sound.  The EoS is then reconstructed as:
\begin{eqnarray}
P(\varepsilon) = \int_0^\varepsilon d\varepsilon'\,c_s^2(\varepsilon')~.
\label{eq:eos}
\end{eqnarray}
Baryon density and chemical potential,
\begin{eqnarray}
n_B = {\partial P\over\partial\mu_B} ~~~~~\text{and} ~~~~\mu_B={\partial\varepsilon\over\partial n_B}~,
\label{eq:nBmuB}
\end{eqnarray}
are determined in accordance with the zero-temperature Gibbs-Duhem equation, 
\begin{eqnarray}
P+\varepsilon = \mu_B\,n_B~.
\label{eq:gibbsduhem}
\end{eqnarray}

\subsubsection{Bayesian inference of the sound speed and of related neutron star properties}

The initial step is to prepare a general parametrization of the squared sound velocity,  $c_s^2(\varepsilon)$.  A convenient choice is a segment-wise representation with a sufficiently large number of segments \cite{Brandes2023, BKW2023}.  This defines a prior ${\cal P}r(\theta)$ in parameter space,  $\theta(c_{s,i}^2, \varepsilon_i)$,  with segments $(i = 1, \dots, N)$ chosen such that the prior maximally covers the available space and also includes the freedom for the possible occurance of phase transitions or crossovers.  

Multiple speed-of-sound parametrizations are then converted into multitudes of trial equations-of-state,  which are in turn confronted with empirical neutron star masses and radii solving Tolman-Oppenheimer-Volkov (TOV) equations.  Given a data set ${\cal D}$,  Bayesian inference is used to compute the posterior ${\cal P}r(\theta|{\cal D})\propto{\cal P}r({\cal D}|\theta)\,{\cal P}r(\theta)$ from the likelihood ${\cal P}r({\cal D}|\theta)$.  It is of prime importance that the posterior must be driven by the data,  not by the choice of the prior ${\cal P}r(\theta)$.  

In some selected cases it is interesting to quantify evidences in terms of Bayes factors.  The evidence for a certain hypothesis $h$ is measured by a Bayes factor ${\cal B}(h) =  {\cal P}r({\cal D}|h)/{\cal P}r({\cal D}|\bar{h})$,  the ratio of likelihoods for hypothesis $h$ versus counter hypothesis $\bar{h}$,  with respect to the given data set ${\cal D}$.

Boundary conditions from perturbative QCD are imposed at asymptotically high energy densities where the sound speed must approach the conformal limit,  $c_s^2 = 1/3$.  PQCD is considered to be applicable at baryon chemical potentials $\mu_B > 2$ GeV corresponding to baryon densities well above 20 times the equlibrium density of nuclear matter, $n_0=0.16$ fm$^{-3}$.  At low densities around $n_B \simeq n_0$,  chiral effective field theory (ChEFT) constraints representing nuclear physics information are commonly imposed.  These ChEFT constraints have frequently been introduced as part of the prior,  whereas we emphasize that they should rather be implemented as a likelihood,  at the same level as the empirical data.  Indeed,  ChEFT as it is usually employed in this context is nothing but a systematic and highly efficient parametrization of hadronic and nuclear physics data fitted in terms of a set of low-energy constants.  However,  using such constraints as a prior,  especially at densities approaching the limits of ChEFT applicability,  would introduce too much of a predetermined bias for the Bayesian inference procedure.  In practice we have instead applied ChEFT constraints (typically at order N3LO) as a likelihood at baryon densities $n_B\lesssim 1.3\,n_0$.

\begin{figure}
\centering
\includegraphics[width=8cm]{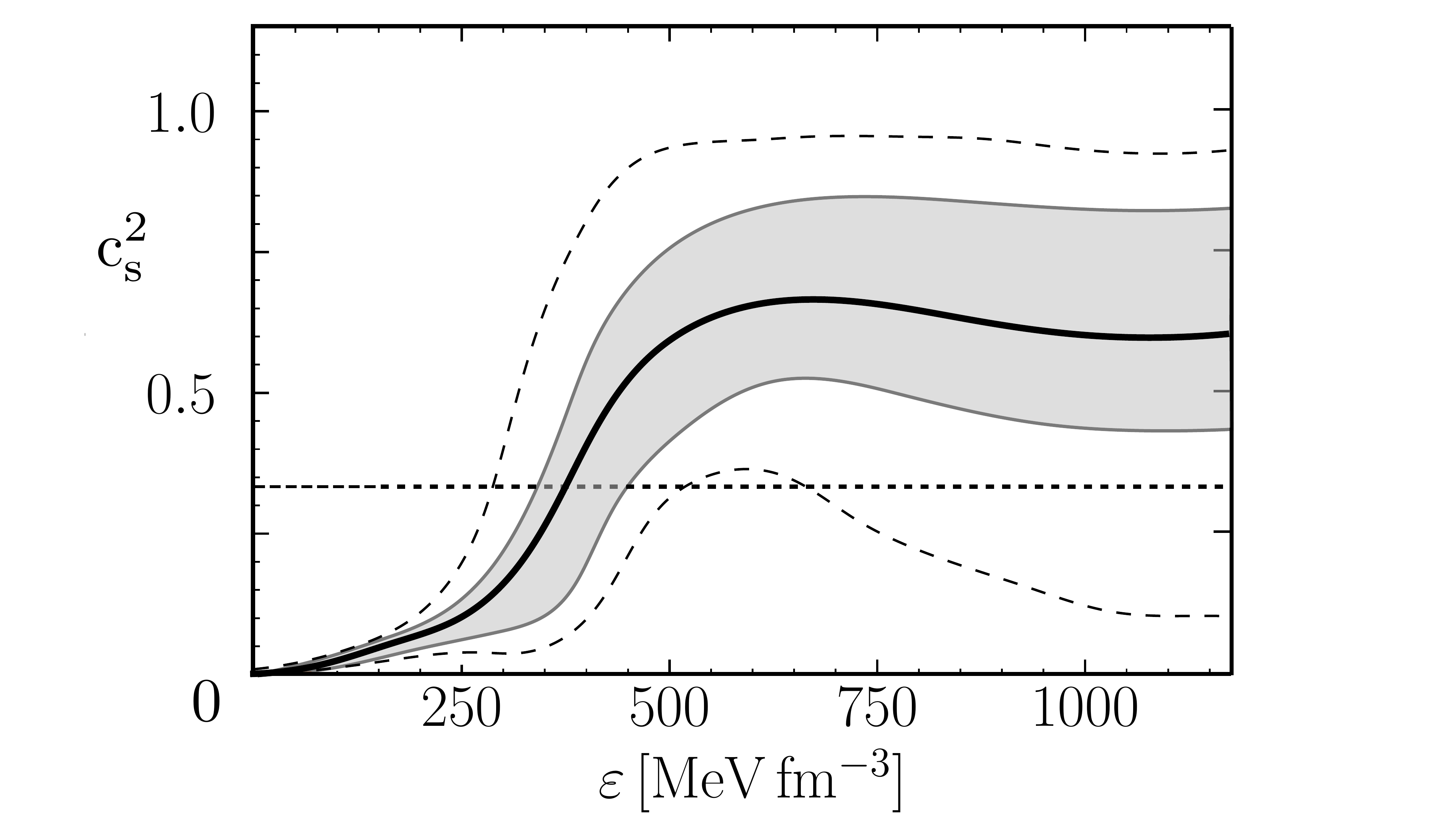}
\includegraphics[width=8cm]{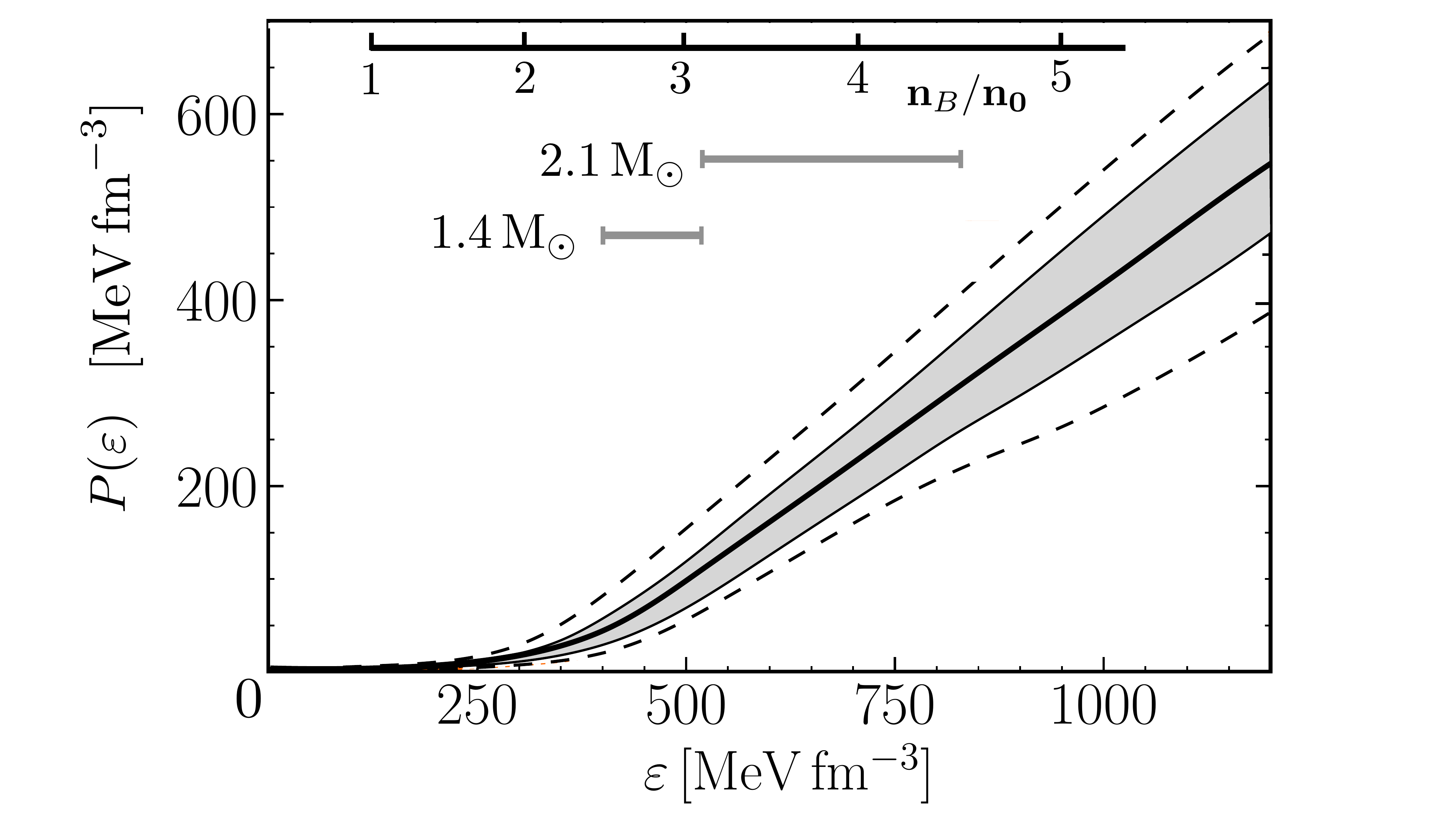}
\caption{Bayes inference results \cite{BW2025} for the squared speed of sound $c_s^2(\varepsilon)$ (upper panel) and pressure $P(\varepsilon)$ (lower panel) of neutron star matter as functions of energy density $\varepsilon$,  using data of Table \ref{tab:Nstardata}.  Medians are shown as solid lines. Grey areas represent 68\% posterior credible bands; dashed lines mark the boundaries of 95\% confidence regions.  The dotted horizontal line in the upper figure indicates the conformal limit,  $c_s^2 = 1/3$.  The baryon density scale $n_B$ in the lower figure (in units of $n_0 = 0.16\,\text{fm}^{-3}$) is derived using the median of $P(\varepsilon)$.  Also shown for orientation are ranges of central densities (at 68\% c.l.) for neutron stars with masses $1.4\,M_\odot$ and $2.1\,M_\odot$,  found by solving TOV equations with the EoS $P(\varepsilon)$. 
\label{fig4} }
\end{figure}

With this input Bayesian inference results of posterior 68\% and 95\% credible bands have been generated for the squared speed of sound and the pressure of neutron star matter as a function of energy density \cite{BW2025} (see Figure\,\ref{fig4}).  The squared sound speed displays a strong increase beyond the conformal limit,  $c_s^2 = 1/3$,  in the range of densities $n_B \simeq 2-3\,n_0$,  until its median saturates at higher densities.  With a Bayes factor well over $10^3$,  there is extreme evidence that $c_s^2 > 1/3$ in the centers of all neutron stars investigated \cite{BKW2023}.  The conformal limit,  characteristic of a relativistic fermi gas,  is quite naturally exceeded in a fermionic many-body system with strongly repulsive correlations.  A correspondingly stiff EoS follows,  capable of supporting the heaviest observed neutron stars against gravitational collapse.

It is instructive solving TOV equations with the inferred $P(\varepsilon)$ as input to reconstruct radial profiles of generic 1.4 $M_\odot$ and 2.1 $M_\odot$ neutron stars.  Their median radii are found to be almost equal at $R \simeq 12$ km,  independent of their mass,  while the 95\% credible band of radii covers the interval $R\simeq 11-13$ km.  The baryon densities at the neutron star centers,  $n_{B,c}(M)$,  turn out not to be extreme.   At 68\%  c.l.  one finds\footnote{These Bayes inferred numbers include the heaviest (`black widow') pulsar PSR J0952-0607 with an equivalent non-rotating mass of $2.3\pm 0.1\,M_\odot$ in the data base.  Omitting this pulsar from the inference analysis raises $n_{B,c}(2.1\,M_\odot)$ to $4.1^{+0.8}_{-0.7}\,n_0$ (at 68\% c.l.) while $n_{B,c}(1.4\,M_\odot)$ is left unchanged.} \cite{BW2025}:
\begin{eqnarray}
n_{B,c}(1.4\,M_\odot) &=& (2.8\pm 0.3)\,n_0~~,\nonumber\\n_{B,c}(2.1\,M_\odot) &=& (3.8^{+0.6}_{-0.7})\,n_0~~.
\label{eq:cenden}
\end{eqnarray}
The central baryon density of even a 2.3 solar mass neutron would not exceed about $5\,n_0$,  corresponding to central energy densities $\varepsilon \lesssim$ 1 GeV/fm$^3$.  It should be emphasized that the EoS is constrained by data only up to that range.  Extrapolations of the EoS beyond that range are possible but have no empirical support. 

\begin{figure}
\centering
\includegraphics[width=8cm]{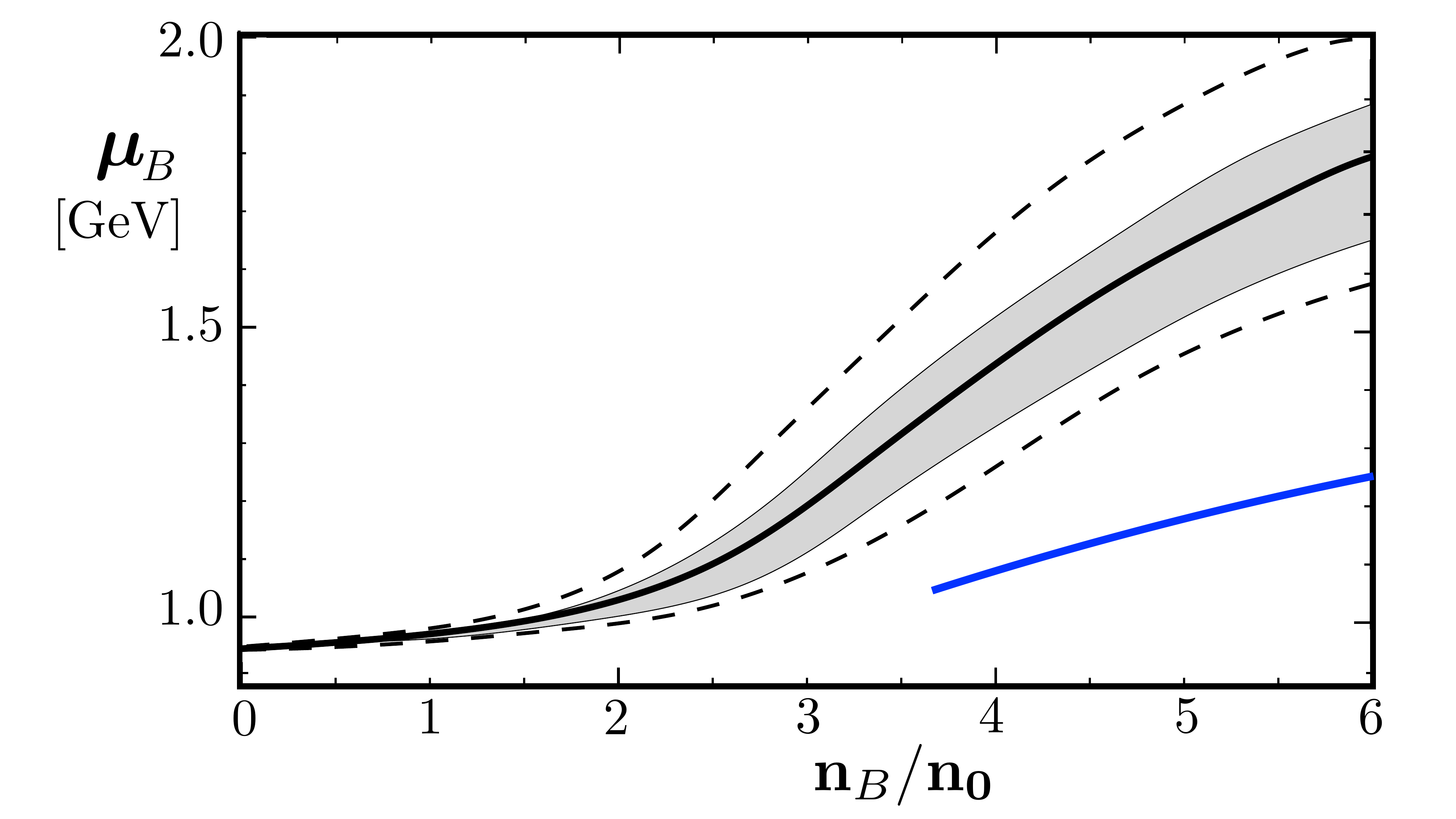}
\includegraphics[width=8.1cm]{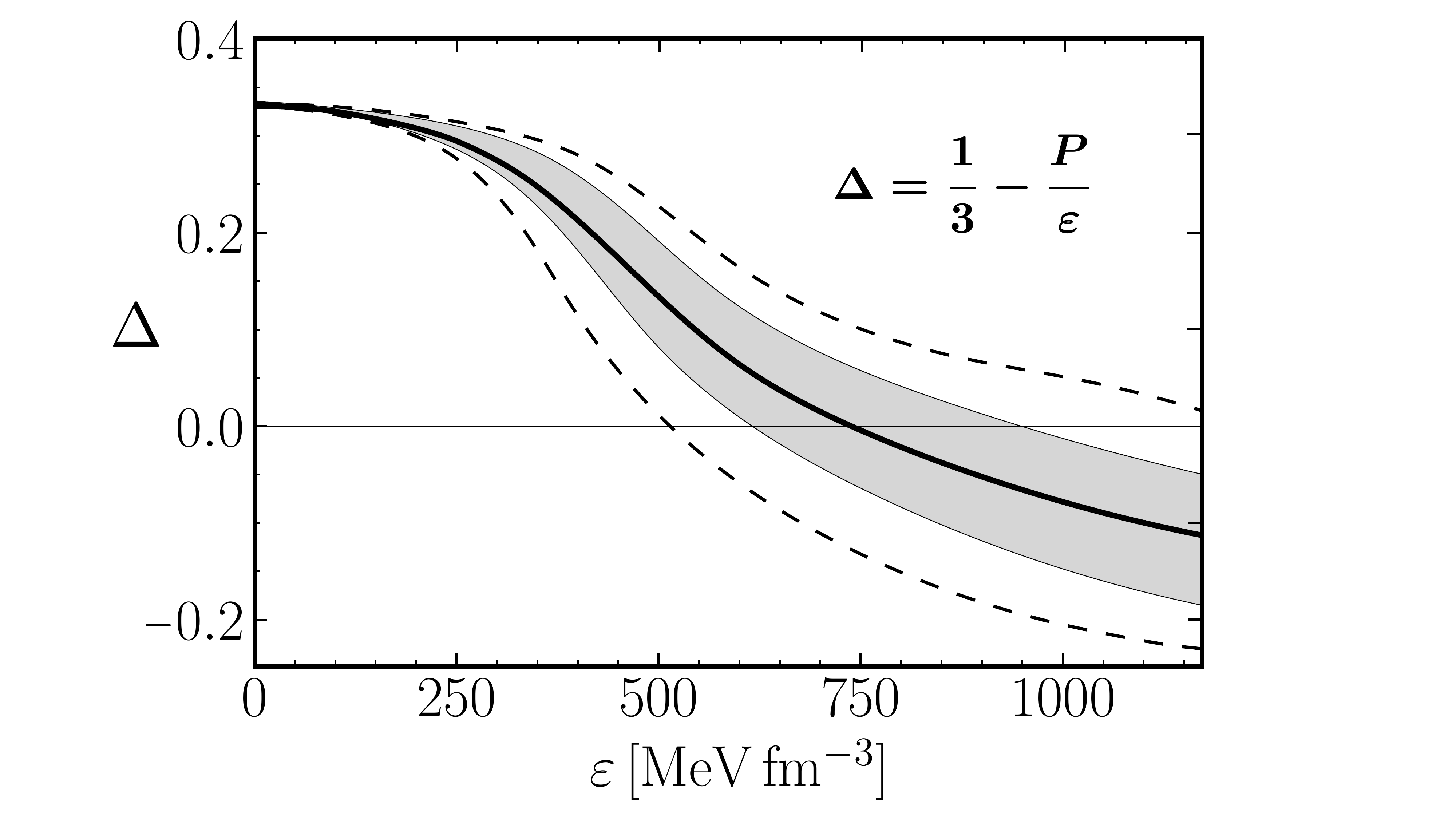}
\caption{Bayes inference results as in Fig.\,\ref{fig4}: Upper panel: baryon chemical potential $\mu_B$ in neutron star matter as function of baryon density $n_B$; shown for comparison (blue curve) is the baryon chemical potential for a quark gas with $N_f=3$ massless flavors.  Lower panel: trace anomaly measure $\Delta = (\varepsilon - 3P)/3\varepsilon$ as function of energy density.  (Adapted from \cite{BW2025}).
\label{fig5} }
\end{figure}

A further quantity of interest is the baryon chemical potential,  $\mu_B$,  as a function of baryon density.  Its distribution can be computed solving Eqs. (\ref{eq:nBmuB},\ref{eq:gibbsduhem}) using the inferred $P(\varepsilon)$.  The result shown in the upper part of Figure\,\ref{fig5} displays again a rapid rise at densities $n_B > 2\,n_0$.  The Bayesian inference procedure cannot provide more detailed information about the fractions of different baryonic species contributing to the total $\mu_B$.  But whichever degrees of freedom (nucleons,  hyperons,  quarks, ...) are actively involved,  there must be strongly repulsive correlations between them to facilitate the empirically contrained $\mu_B$.  

An instructive guide for orientation is the baryon chemical potential of a gas of massless quarks with $N_f = 3$ flavors.  As demonstrated in Figure\,\ref{fig5},  the assumption of non-interacting quark matter in the center of neutron stars would fail.  Strongly repulsive interactions between quarks are mandatory,  as introduced for example in a model featuring a continuous hadrons-to-quarks crossover (QHC21) \cite{Kojo2022}.  The baryon chemical potential of this model compares favourably with the empirically constrained one.

Much discussion has been focused on the question whether the conformal limit in QCD,  $c_s^2\rightarrow 1/3$, could possibly be reached already at densities accessible in the centers of heavy neutron stars.  The quantity of interest is the trace anomaly or conformality measure \cite{Fujimoto2022}:
\begin{eqnarray}
\Delta ={T_\mu^\mu\over 3\varepsilon} = {1\over 3} - {P(\varepsilon)\over\varepsilon}~,
\label{eq:traceanom}
\end{eqnarray}
where $T_\mu^\mu$ is the trace of the QCD energy momentum tensor.  Using again the inferred $P(\varepsilon)$ posterior distributions,  the result for $\Delta$ is shown in the lower part of Figure\,\ref{fig5}.  Conformality $(\Delta =0)$ is not yet reached at the maximum energy densities chracteristic of neutron stars.  In fact a detailed Bayes factor analysis provides strong evidence for negative $\Delta$ at baryon densities $n_B \simeq 4-6\,n_0$: the pressure exceeds one third of the energy density in that density range and thus maintains the required stiffness of the EoS.

\subsubsection{Comparisons with alternative approaches}

\begin{figure}
\centering
\includegraphics[width=8.5cm]{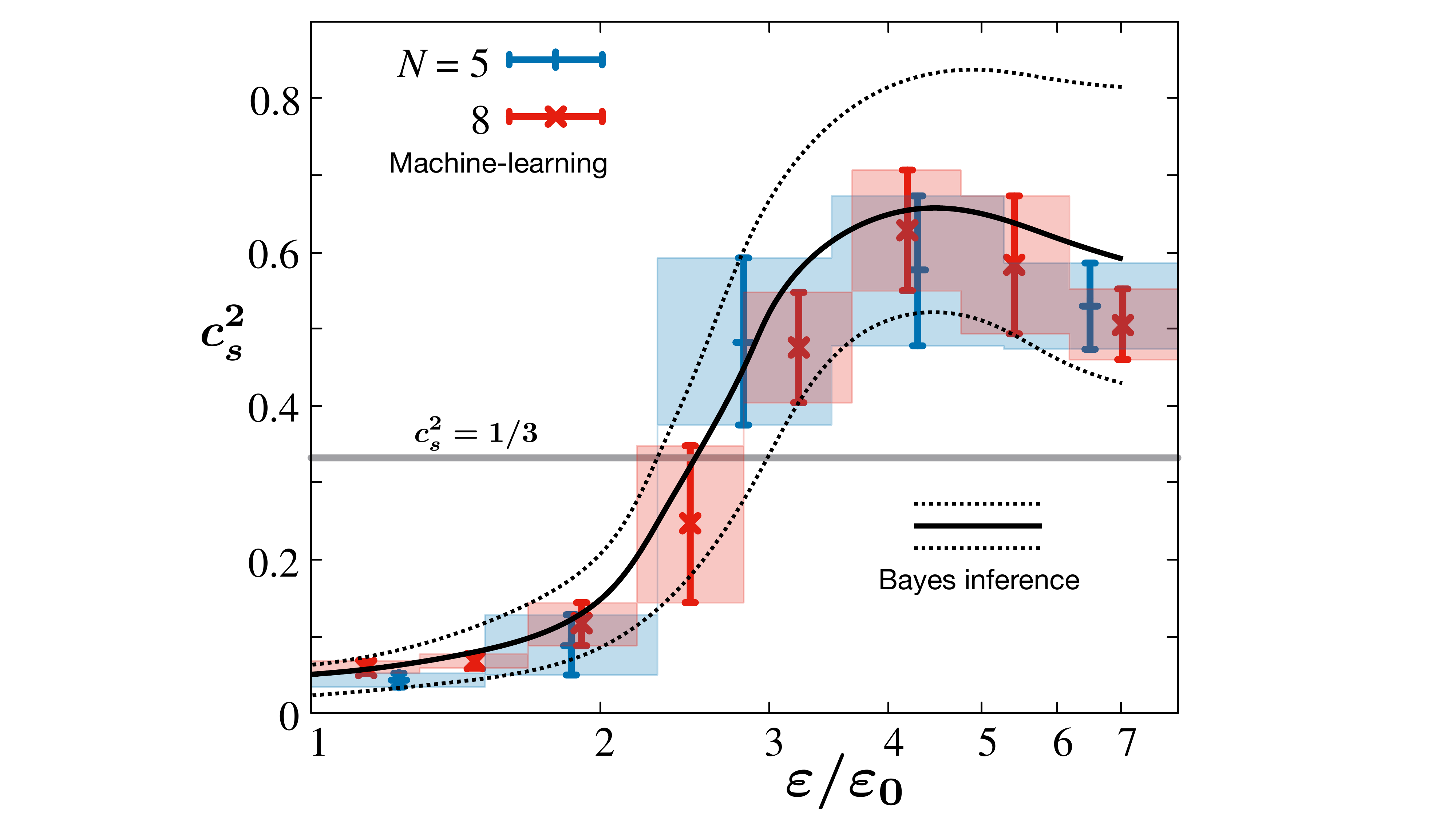}
\caption{Comparison of two independent inference results for the squared speed of sound, $c_s^2(\varepsilon)$, of neutron star matter:  machine-learning approach using $N$ segments of polytropes \cite{Fujimoto2024} vs.  Bayes inference posterior (median and 68\% credible band) \cite{BW2025}.  The energy density on the horizontal axis is plotted in units of $\varepsilon_0 = M_N n_0 = 0.15$ GeV/fm$^3$.
\label{fig6} }
\end{figure}

It is a useful exercise to check the Bayesian inference results described  in previous sections against other independent approaches reported in the literature.   A good example is the inference of the speed of sound using machine learning (neutral network) methods performed in \cite{Fujimoto2024}.  Figure\,\ref{fig6} presents the comparison of the Bayes inferred median and 68\% credibility band with the machine-learning inferred results for $c_s^2$ including uncertainty estimates.  Evidently the test of consistency between these two basically different approaches using essentially the same empirical data sets turns out quite successful.  

\begin{figure}
\centering
\includegraphics[width=9.5cm]{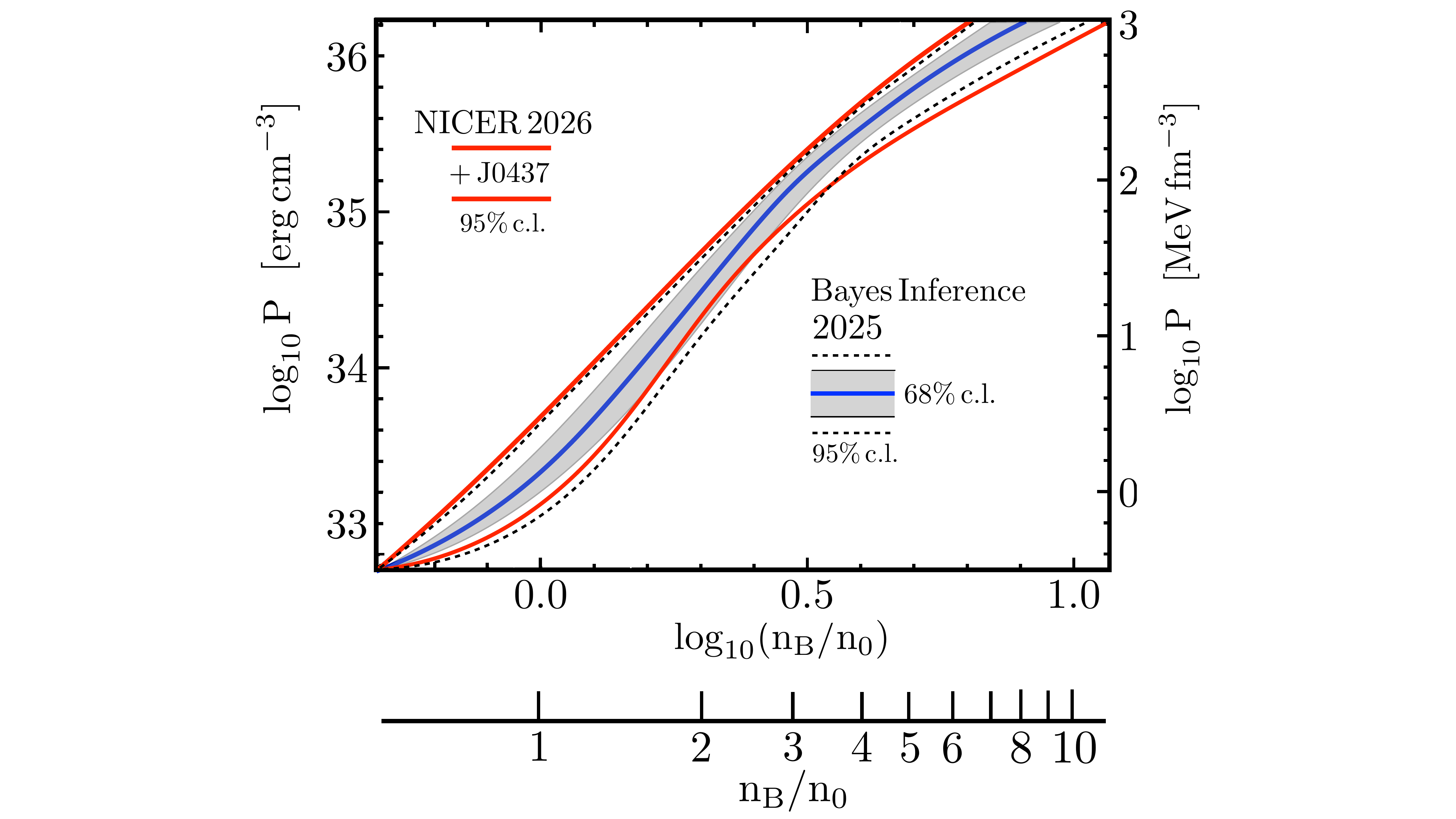}
\caption{Consistency test of neutron star matter equations of state: EoS inferred from NICER data including updated PSR J0437-4715 analysis \cite{Miller2026} (red lines enlosing 95\% c.l.  area),  compared to Bayes inference 65\% and 95\% c.l.  posterior bands \cite{BW2025} (as in Fig.\,\ref{fig4}).
\label{fig7} }
\end{figure}

Another test case of interest concerns the sensitivity of the inferred EoS with respect to changes when improved data analysis leads to updated values e.g.  of neutron star radii.  An example of this kind is the 1.4 $M_\odot$ pulsar PSR J0437-4715 for which the radius reported by NICER has shifted from previous $R\simeq 11.4$ km (with about 10\% uncertainty) to the updated $R$ value listed in Table\,\ref{tab:Nstardata}.  As pointed out in \cite{Miller2026},  this shift of the radius does not change conclusions about the EoS in any significant way.  It is instructive to compare the inferred pressure bands,  $P(n_B)$ as function of baryon density,  from the NICER analysis \cite{Miller2026}, with the Bayes-inferred ones \cite{BW2025},  as displayed in Figure\,\ref{fig7}.  The degree of consistency between these two independent results is quite satisfactory. 

\subsubsection{Constraints on phase transitions in neutron star matter}

Continuing discussions deal with the possible existence of exotic objects such as hybrids (neutron stars in which hadron and quark phases coexist) or twin stars (a pair of neutron stars with degenerate masses but different radii).  Each one of these objects could be indicative of a first-order phase transition in the EoS.  It is thus interesting to examine to what extent the data-driven Bayes inferred EoS can set constraints on the occurence of such phenomena \cite{BKW2023, Brandes2024}.

\begin{figure}
\centering
\includegraphics[width=9cm]{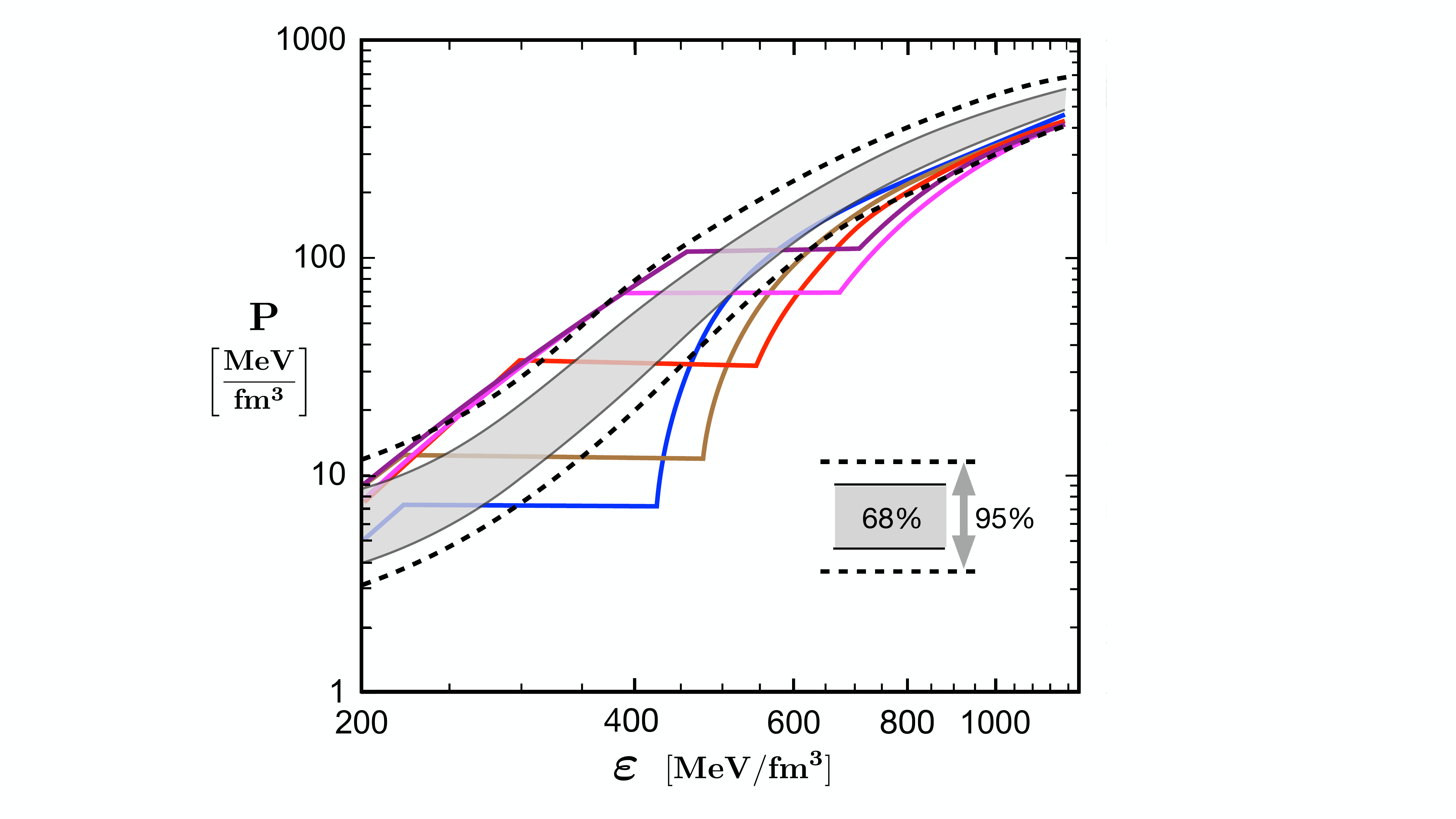}
\caption{Illustration of how the data-driven Bayes-inferred EoS of neutron star matter \cite{BW2025} sets tight constraints for possible scenarios involving strong first-order phase transitions \cite{Li2024}.  Grey area and dashed border lines show 68\% and 95\% posterior credible bands as in Fig.\,\ref{fig4}.
\label{fig8} }
\end{figure}

Numerous studies exploring the possible relevance of first-order phase transitions for neutron star matter have appeared in the literature.  A representative example \cite{Li2024} discusses the possiblity of strong first-order transitions featuring broad hadron-quark coexistence regions of constant pressure.  Figure\,\ref{fig8} shows a confrontation of several such cases with the empirically constrained EoS,  i.e.  the inferred posterior bands of the pressure $P(\varepsilon)$ \cite{BW2025}.  Evidently none of the depicted 1st order transition scenarios is compatible with the $95\%$ credibility band of the EoS.  Some of these transitions appear at very low baryon densities not far above $n_0$,  contrasting established nuclear phenomenology\footnote{As a reminder,  the only first-order phase transition known in nuclear physics is the liquid-gas transition in nuclear matter \cite{KW2026}.  It  disappears however in neutron-rich matter when the proton fraction approaches $x_p < 0.1$.}.

Detailed Bayes factor studies support the expectation that a strong first-order phase transition is not likely in neutron star cores,  rendering twin-star solutions highly improbable \cite{Blomqvist2025}.  Indeed there is extreme evidence (with a Bayes factor well above $10^2$) \cite{BKW2023} excluding a strong first-order transition with Maxwell construction for neutron stars with masses $M \lesssim 2\,M_\odot$.

\subsection{Cold dense baryonic matter: \\distance scales and possible scenarios} 

The constraints on the EoS and on related properties of matter in neutron star cores are based entirely on empirical data.  As such they do not offer deeper insights into the more detailed composition of this highly compressed many-body system.  It is nonetheless instructive at this point to recall the maximum range of baryon densities (\ref{eq:cenden}) that can be reached at the centers of neutron stars when their profiles are constructed solving TOV equations with sets of constrained EoS's.  As pointed out,  even for the heaviest known neutron stars these central densities turn out to be limited to $n_B\lesssim 5\,n_0$ (at 68\% c.\,l.).  It is interesting to examine what this implies for the average distances between baryonic constituents under such conditions.  

\begin{figure}
\centering
\includegraphics[width=8cm]{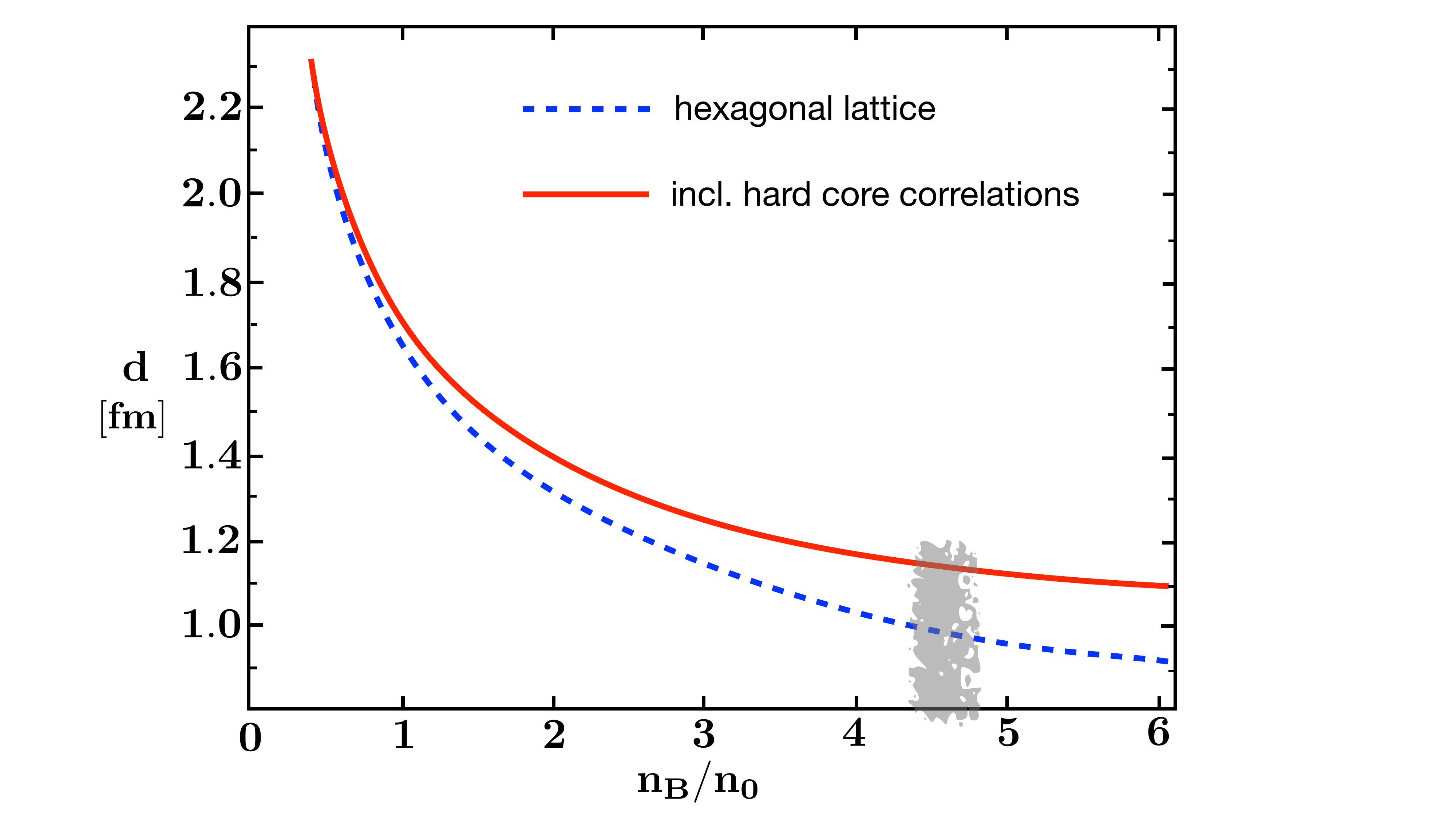}
\caption{Average distance $d(n_B)$ of baryons arranged in a hexagonal lattice as function of density $n_B$ expressed  in units of nuclear matter equlibrium density $n_0$.  Solid curve: with corrections from excluded volume effects.  Dashed curve: no such corrections.  In both cases parameters are fine-tuned to reproduce $d(n_B=n_0) \simeq 1.7$ fm.  The shaded grey area indicates the range of maximal central densities that can be reached in the most massive neutron stars.
\label{fig9} }
\end{figure}

For a quick estimate,  imagine a dense system of baryons closely packed in a hexagonal lattice arrangement (with packing fraction $\phi = {\pi\over 3\sqrt{2}} =  0.74$).  Adding an excluded-volume correction $v_x$ to simulate short-range repulsive correlations,  the average distance between baryons at a density $n_B$ is $d(n_B) = [\phi/n_B(1-v_x\,n_B)]^{1/3}$.  Fixing $v_x$ such that the average distance between two nucleons in nuclear matter in equlibrium at $n_B = n_0 = 0.16\,\text{fm}^{-3}$ is reproduced at $d(n_0) = 1.7$ fm,  one arrives at the solid curve in Figure\,\ref{fig9}.  Note that even at $n_B \sim 5\,n_0$ the average distance between baryons in such a system is still larger than 1 fm.

This finally leads back to the two-scales picture of the nucleon,  inspired by chiral symmetry of QCD,  and worked out in Section \ref{sec:sizes}.  It suggests a possible scenario for dense matter in systems such as neutron star interiors,  at least in a baryon-dominated description.  As previously outlined,  there are strong and quantifiable indications that the nucleon contains a compact core with an r.m.s. radius of about 1/2 fm which hosts the valence quarks,  their baryon number distribution and most of the nucleon's mass.  This core is surrounded by a soft surface of quark-antiquark pairs,  the mesonic cloud,  governed by pions and their characteristic Nambu-Goldstone boson nature.  As also mentioned,  in dense matter one expects that the mesonic cloud expands following the rules of spontaneously broken chiral symmetry and the density dependence of its order parameter.  The compact cores,  on the other hand,  are supposed to remain quite stable up to densities when they begin to touch and overlap.  

\begin{figure}
\centering
\includegraphics[width=8cm]{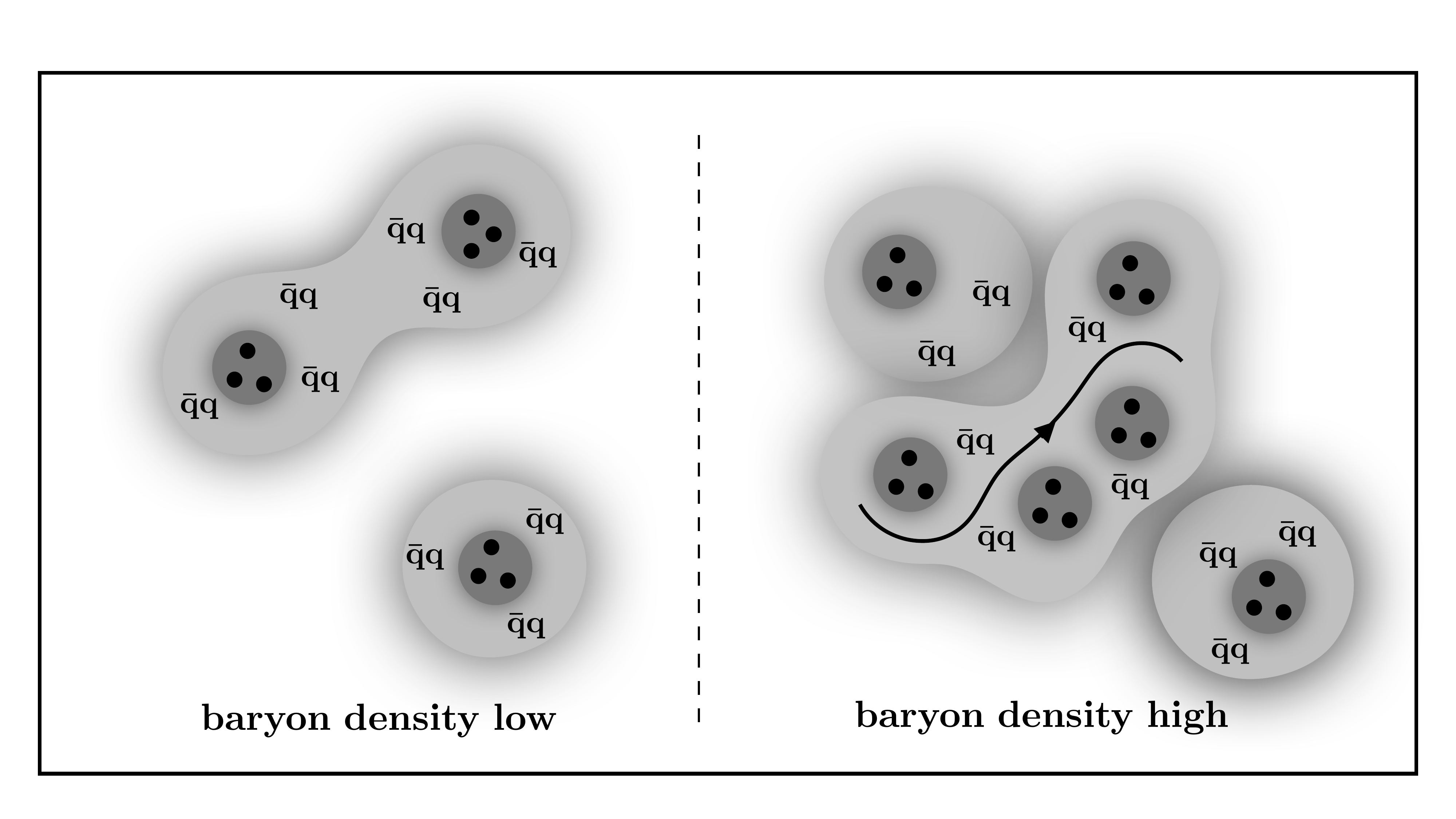}
\caption{Illustration of a scenario for low- and high-density baryonic matter.  Baryons (e.g. nucleons) are viewed in a two-scales `core-plus-cloud' picture based on chiral symmetry of QCD: compact valence quark cores enclosing the baryon number distribution and most of the nucleon mass,  surrounded by soft quark-antiquark clouds dominated by pions.  Left: low density with isolated nucleons and occasional two-body interactions.  Right: high density with overlapping meson clouds (percolating $\bar{q}q$ pairs) generating many-body interactions.  (Figure adapted from \cite{Brandes2024}).
\label{fig10} }
\end{figure}

A two-scales scenario of this kind for dense baryonic matter has been proposed in \cite{Fukushima2020}.  Figure\,\ref{fig10} presents an illustrative sketch\footnote{Mannque appreciated such a picture and took it over into a publication \cite{Rho2024} where he gave an extensive account of the `Cheshire Cat' principle that he promoted. } of the emerging physics as it can be conceived in the pertinent density regimes:
\begin{itemize}
\item{At densities $n_B \lesssim n_0 = 0.16\,\text{fm}^{-3}$ and average distances $d\gtrsim 1.5$ fm the baryons and their cores are well separated.  Tails of the mesonic clouds overlap occasionally to form two-body boson exchange forces between the baryons.}
\item{At densities $n_B \gtrsim 2 - 3\,n_0$ (distances $d \sim 1.2-1.4$ fm) the soft mesonic clouds begin to delocalize.  The mobility of $\bar{q}q$ pairs increases as percolation sets in and the mesonic fields cover more than two baryons simultaneously.  Many-body forces develop.  The baryonic cores are still separated but experience increasingly strong repulsive Pauli effects as the density rises.}
\item{At densities $n_B > 5\,n_0$ (beyond the central core densities that can be reached in even the heaviest neutron stars) the compact baryonic valence quark cores begin to touch and overlap at distances $d \lesssim 1$ fm.  Upon further compression they still have to overcome the short-distance repulsive core characteristic of the nucleon-nucleon interaction.}
\end{itemize}

\section{Epilogue}

A scheme of two underlying scales such as the one just described puts into perspective the notion of deconfinement that originated in early versions of the quark bag model with its large bag radius of about 1 fm. This would have implied that the deconfinement of valence quarks starts already around densities $n_B\sim n_0$,  in contradiction with well established nuclear physics phenomenology.  But in fact there is no sharp bag boundary that could serve to define a measurable confinement radius.  {\it Chiral} bag models had already challenged the idea that a confining bag boundary condition could have a physically observable meaning.  Mannque Rho emphasized this again in his recent review \cite{Rho2024}.  

With increasing density in baryonic matter one expects first a continuously soft transition featuring a delocalization of the quark-antiquark clouds attached to the baryonic surfaces.  This mechanism can be viewed as a percolation process in which the mesonic fields progressively cover an increasing number of still separated valence quark clusters with their non-overlapping baryon number distributions \cite{Fukushima2020}.  The language of chiral effective field theory \cite{Epelbaum2009, Drischler2021} describes such a process systematically in terms of an expanding hierarchy of interactions,  organized as a series in powers of Fermi momentum,  from familiar Yukawa type two-body potentials at low densities to many-body forces at higher densities.

In the two-scales core-plus-cloud picture of the nucleon,  a `deconfinement' of valence quarks would be anticipated only at baryon densities well above $5\,n_0$,  exceeding the expected maximum central densities in neutron stars.  
For that to take place the compact baryonic cores must first overcome their strongly repulsive interactions at distances well below 1 fm until their baryon number distributions overlap and release quarks into the continuum,  along with a restoration of chiral symmetry in its Wigner-Weyl mode.  As argued in \cite{Brandes2021},  this would likely proceed not as a phase transition but as a continuous crossover. 

\subsection{Acknowledgements}

Feelings of deep gratitude and lively memories are devoted to Mannque Rho and the numerous discussions we enjoyed together over many decades.

Results presented in several sections of this manuscript were obtained in close collaborations with Len Brandes and Norbert Kaiser for which I am very grateful.  I thank Alex Dittmann for much useful first-hand information about NICER data and their analysis.  I also thank Kenji Fukushima,  Avraham Gal,  Bao-An Li,  Ulf-G.  Mei{\ss}ner,  Krzysztof Redlich,  Chihiro Sasaki and Ismail Zahed for recent stimulating exchanges. 

This work has been partially supported by the German Research Foundation (DFG) under EXC-2094/2-390783311 (Exc. Cluster ORIGINS).

\end{document}